# Electron Density and Temperature in NIO1 RF Source Operated in Oxygen and Argon

M. Barbisan[1, a)], B. Zaniol[1], M. Cavenago[2], R. Pasqualotto[1], G. Serianni[1], M. Zanini[3]

[1] *Consorzio RFX (CNR, ENEA, INFN, University of Padova, Acciaierie Venete SpA), C.so Stati Uniti 4 – 35127, Padova (Italy).*
[2] *INFN-LNL, v.le dell'Università 2, I-35020, Legnaro (Italy)*
[3] *Università degli Studi di Padova, via 8 Febbraio, 2 – 35122 Padova*

[a)] Corresponding author: marco.barbisan@igi.cnr.it

**Abstract.** The NIO1 experiment, built and operated at Consorzio RFX, hosts an RF negative ion source, from which it is possible to produce a beam of maximum 130 mA in $H^-$ ions, accelerated up to 60 kV. For the preliminary tests of the extraction system the source has been operated in oxygen, whose high electronegativity allows to reach useful levels of extracted beam current. The efficiency of negative ions extraction is strongly influenced by the electron density and temperature close to the Plasma Grid, i.e. the grid of the acceleration system which faces the source. To support the tests, these parameters have been measured by means of the Optical Emission Spectroscopy diagnostic. This technique has involved the use of an oxygen-argon mixture to produce the plasma in the source. The intensities of specific Ar I and Ar II lines have been measured along lines of sight close to the Plasma Grid, and have been interpreted with the ADAS package to get the desired information. This work will describe the diagnostic hardware, the analysis method and the measured values of electron density and temperature, as function of the main source parameters (RF power, pressure, bias voltage and magnetic filter field). The main results show that not only electron density but also electron temperature increase with RF power; both decrease with increasing magnetic filter field. Variations of source pressure and plasma grid bias voltage appear to affect only electron temperature and electron density, respectively.

## INTRODUCTION

Optimizing the production of $H^-/D^-$ ion beams is a crucial task for the operation and the efficiency of future fusion reactors. To contribute to this purpose Consorzio RFX and INFN-LNL have designed, built and operated the NIO1 (Negative Ion Optimization 1) experiment [1-3]. NIO1 hosts a small (10 cm diameter x 21 cm length) and flexible negative ion source, in which a plasma is created thanks to the RF power (max. 2.5 kW) radiated by a solenoid. No electrostatic shields are present, however most inner surfaces (except those between the plasma and the solenoid) are protected by a 0.5 mm thick Mo foil. The negative ions produced in the plasma can be extracted and accelerated by a set of grids, composed by: Plasma Grid (PG-facing the source), Extraction Grid (EG), Post Acceleration Grid (PA-at ground voltage) and Repeller (REP). Source and acceleration system are shown in fig. 1a. At its best performance, NIO1 is designed to produce a 130 mA beam of $H^-$ ions accelerated at 60 keV, in continuous (>1000 s) operation.

The properties of the plasma in the source can be modified by electric and magnetic fields in a very flexible way [4]; the main aims are

- to keep the electron temperature low, in order to avoid the electron stripping reactions undergone by negative ions because of electron impact;
- to limit the amount of electrons co-extracted from the source.

The PG can be biased positively with respect to the source body, so that electrons hit the PG surfaces instead of passing through the grid holes. A separate voltage can be applied also between the PG and the electric bias plate

($BP_e$, shown in fig. 1b), which surrounds the PG apertures at a distance of 4÷14.5 mm. To lower the electron temperature, a magnetic field is generated by a current $I_{PG}$ (max. 400 A) flowing through the PG. The intensity and the extension of the magnetic field can be varied not only by changing the current intensity but also by changing the components of the source through which the $I_{PG}$ circuit is closed. The PG current can be brought back to the power supply through the magnetic Bias Plate ($BP_m$, shown in fig. 1b), a rectangular tube surrounding the PG apertures at a distance of 9 mm. Alternatively, the PG current can flow back through 2 "C" coils (shown in fig. 1a), i.e. 2 semicircle-shaped external wires on the top and bottom halves of the source. Being the "C" coils at 60 mm from the PG, the produced magnetic field is more extended in space with respect to the circuit with the $BP_m$.

The flexibility in the source configuration is reflected on the behavior of the plasma and then on the properties of the produced beam. Currently, the only way to monitor the source plasma is by analyzing the light emitted by it. In particular, NIO1 is equipped with an Optical Emission Spectroscopy diagnostic (OES), mainly from prototypes of the SPIDER OES diagnostic [5], which allows to get the spectrum of the light emitted from the plasma. The interpretation of the measured spectral lines has been used to measure important quantities of the plasma.

The study presented in this paper will show the main measurements obtained by the OES diagnostic on NIO1, during the experimental campaigns in which hydrogen was replaced by a mixture of oxygen (90%) and argon (10 %) [6]. This mixture was preferred to hydrogen to perform the tests for the acceleration system; oxygen is indeed more electronegative than hydrogen, therefore it is possible to extract a higher current of negative ions from the source. Argon was introduced for diagnostic purposes: from the analysis of Ar I and Ar II lines it was possible to get the electron temperature and density of the plasma, thanks to the data made available by the ADAS software package [7]. Of course, a Maxwellian distribution of electrons had to be assumed. The measurements of electron density and temperature were helpful to assess the effectiveness of the methods used to limit the co-extracted electrons and the destruction rate of negative ions by electron impact.

The paper will present the instrumentation of the OES diagnostic and the model used to interpret the collected spectra. The paper will then discuss how electron density and temperature, according to the OES measurements, are affected by RF power, source pressure, PG bias voltage and magnetic filter field.

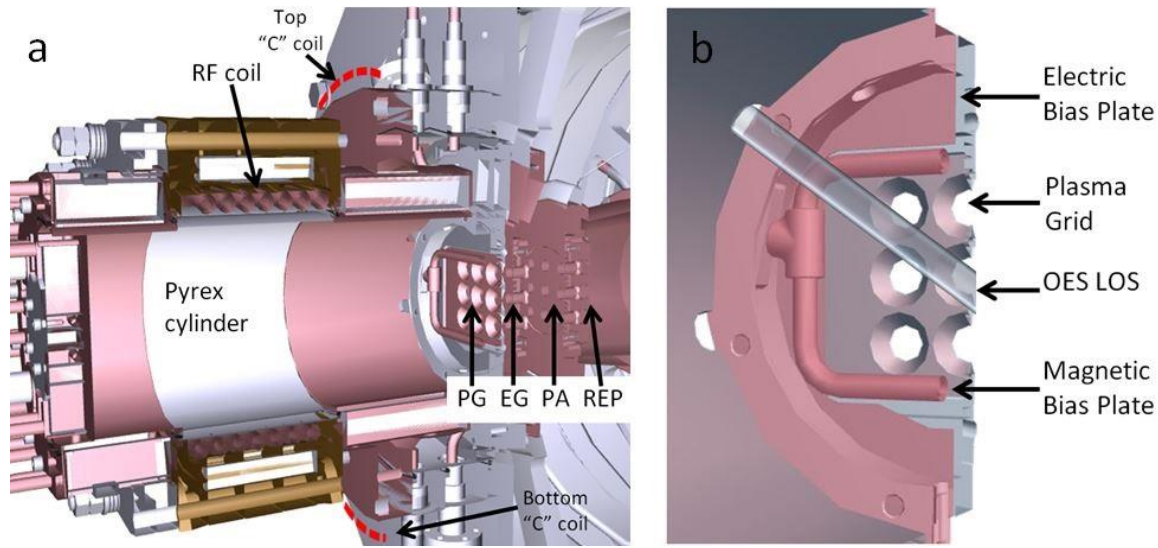

**FIGURE 1.** a) 3D section of the NIO1 source and its acceleration system. b) 3D section of the source components close to the PG, together with the volume of the LOS dedicated to the OES diagnostic.

## THE OES DIAGNOSTIC IN NIO1

The OES diagnostic [9] has access to the light emitted inside the source through viewports with clear apertures of 8 mm diameter. The light exits through fused silica windows, no significant attenuation is expected at the wavelengths of the lines used in the analysis. The light is collected along 2 Lines of Sight (LOSs), by means of optic heads equipped with plano-convex BK7 lenses of 50 mm focal length and 4 mm clear aperture. The lenses convey the light on silica-silica optical fibers with 400 µm core diameter. One of the 2 fibers is connected to a low resolution spectrometer, namely a *Hamamatsu C10082CAH* [10], mounting an integrated back thinned CCD sensor of 2048 pixels. The device has a plate factor of 0.30÷0.36 nm/pix and a resolution of 1 nm; acquired spectra range from 350 to 850 nm. The low resolution spectrometer has been mainly used for a qualitative survey of the plasma conditions and for the identification of possible impurities in the plasma. The other fiber is instead connected to a high resolution spectrometer, namely an *Acton SpectraPro-750* [11], equipped with a 2D back illuminated frame transfer CCD camera of 512x512 pixel, for 6 nm wide spectral window, a plate factor of 0.012 nm/pix and a resolution of 50 pm (entrance slit width set to 50 µm). The high resolution spectrometer was used to resolve the argon lines which were necessary to get the electron temperature and density. Both the spectrometers, together with fibers and related optic heads, have been absolutely calibrated with an Ulbricht sphere in order to calculate the radiance of the observed lines, or to simply correct their relative intensity.

The two optic heads are mounted on viewports that are at opposite positions and directions. The common LOS is parallel to the PG, at a distance of 19 mm from its surface; as shown in fig. 1a and 1b, the LOS is just upstream with respect to $BP_m$ and $BP_e$, but downstream with respect to the "C" coils.

## THE INTERPRETATION OF OXYGEN-ARGON SPECTRA

The spectra acquired by the OES diagnostic show a variety of atomic lines and molecular bands. An example is given in fig. 2, showing the high resolution spectrum (black curve) of the plasma light in the wavelength interval 350÷850 nm. The plasma in the source was sustained by 800 W RF power at a gas pressure of 0.48 Pa. The spectrum is expressed in emissivity units, photons/(m$^3$·s·nm), taking into account the absolute calibration curve and assuming that the plasma is uniform over the 10 cm diameter of the source. Fig. 2 also reports an example of low resolution spectrum (red curve), again in emissivity units (red axes on the right), which was acquired under similar operative conditions with respect to the high resolution one. The different levels of emissivity are due to the different widths of the instrumental functions of the 2 spectrometers.

In the OES spectra many lines of atomic oxygen can be detected, in particular the triplet at 777 nm (saturated in the plots) is an order of magnitude more intense than all other lines. No lines of O$^+$ have been detected, while various molecular bands of the first ($b^4\Sigma_g^- \rightarrow a^4\Pi_u$) and second ($A^2\Pi_u \rightarrow X^2\Pi_g$) negative systems of O$_2^+$ have been found [12]. Besides oxygen, many lines of atomic and singly ionized argon have been detected.

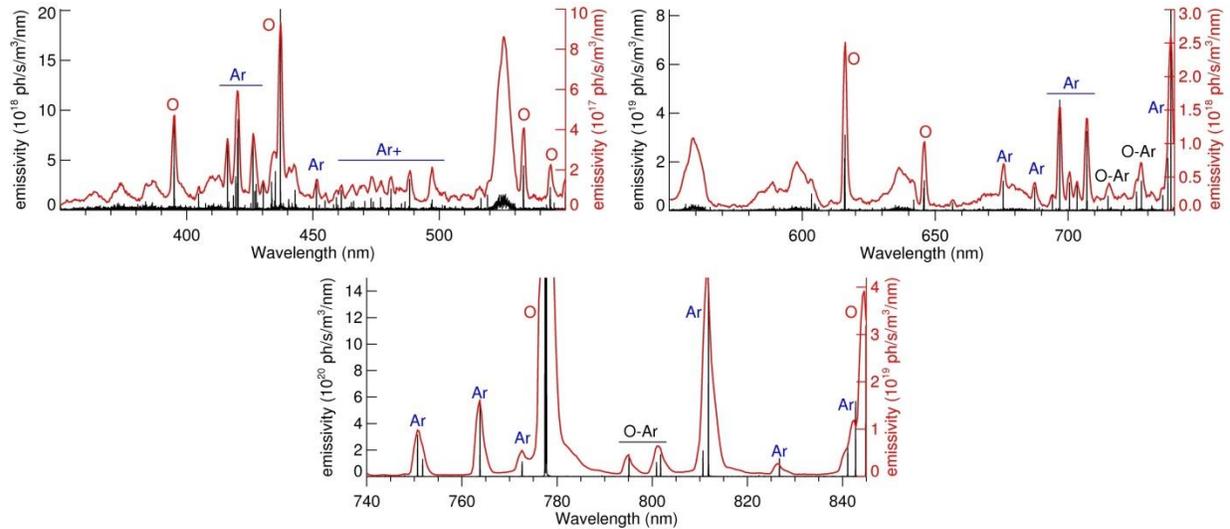

**FIGURE 2.** Absolutely calibrated spectrum (black curve) acquired by the high resolution spectrometer, for a plasma at a pressure of 0.48 Pa and sustained by 800 W RF power. About 80 spectra have been used to reconstruct the whole high resolution spectrum. The main atomic lines of oxygen, argon and singly ionized argon are indicated. The plots also show a low resolution spectrum, acquired under similar plasma conditions; the spectrum is represented with a red curve, again in emissivity units (y axes on the right).

To measure the electron temperature and the electron density the Ar II (i.e. Ar$^+$) lines at 480.6 nm and 488.0 nm and the Ar line at 750.4 nm were exploited. As fig. 2 suggests, resolving these lines from those at nearby wavelengths was possible only by means of the high resolution spectrometer. The emissivities of the selected lines ($\varepsilon_{480}$, $\varepsilon_{488}$, $\varepsilon_{750}$) were obtained from the spectra by means of Gaussian fits, and were interpreted by means of the ADAS [7-8] Collisional-Radiative (CR) model for argon. The CR model gives the effective emission rates ($X_{480}$, $X_{488}$ and $X_{750}$) of the 3 considered lines, as function of the electron density $n_e$ and of the electron temperature $T_e$. The relation between the emissivities and the effective emission rates is the following:

$$\varepsilon_{480} = n_{Ar+} n_e X_{480}(n_e, T_e) \quad \varepsilon_{488} = n_{Ar+} n_e X_{488}(n_e, T_e) \quad \varepsilon_{750} = n_{Ar} n_e X_{750}(n_e, T_e) \tag{1}$$

where $n_{Ar}$ and $n_{Ar+}$ are the densities of atomic and singly ionized argon.
The electron temperature and the electron density were calculated with an iterative procedure:
- The electron temperature was initially set to a guess value.
- The electron density was then measured from the line emissivity ratio $\varepsilon_{480}/\varepsilon_{488}$, which is equal to $X_{480}/X_{488}$. For electron temperatures above 1 eV this ratio mainly depends on the electron density; the values of $X_{480}/X_{488}$ calculated from the CR model were interpolated for the guess value of $T_e$ and then compared to the experimental value of $X_{480}/X_{488}$ to retrieve an estimate of $n_e$.
- The argon density was calculated as 10 % of the gas density, in turn estimated from the gas pressure and the gas temperature. This was assumed to be 300°K. In principle it should be possible to estimate the gas temperature from the atmospheric A band of $O_2$ ($b^1\Sigma_g^+$, $v=0 \rightarrow X^3\Sigma_g^-$, $v=0$) at 760 nm [13]; the lines of the band were however too low to be detected.
- At this point the electron temperature was measured by exploiting the argon line at 750 nm: the experimental value of $X_{750}$ was obtained from the emissivity $\varepsilon_{750}$ of the line and from the values of $n_e$ and $n_{Ar}$, estimated in the previous steps. The values of $X_{750}$ output by the CR model for different values of $n_e$ and $T_e$ were interpolated for the estimate of $n_e$; the resulting set of $X_{750}$ values was then compared to experimental value of $X_{750}$ to get the electron temperature.
- The new value of $T_e$ replaced the initial guess and the whole calculation was repeated a number of times sufficient to get an adequate convergence of the results.

## RESULTS IN VARIOUS OPERATIVE CONDITIONS

The aforementioned analysis technique was exploited to investigate how electron temperature and electron density depend on various operative parameters and source electrical configurations.

Figure 3 shows the measurements of electron temperature (plot a) and electron density (plot b), performed with the source operated at 0.48 Pa gas pressure and at RF power levels between 400 W and 1000 W. The measurement at each RF power value was repeated for different values of PG current ($I_{PG}$) i.e. of magnetic filter field. The source was configured so that $I_{PG}$ flows through the PG and the "C" coils; $BP_m$ and $BP_e$ were instead connected and left at floating potential. In this configuration, the magnetic field generated at the position of the LOS is about 75 G/$I_{PG}$[kA]. Fig. 3b shows that the electron density, as expected, increases with RF power.

According to a simplified model [14], from the particle balance between ions produced by electron impact and ions lost on the source walls, it is expected that the electron temperature is independent of the electron density. A higher income in RF power should then lead to an increase in electron density, but not in electron temperature. $T_e$ should instead grow if the density of the neutral particles decreases.

What experimentally results in the case of NIO1 is that the electron density, as expected, increases with RF power (fig. 3b). Contrarily to what expected, instead, also the electron temperature grows with increasing RF power (fig. 3a), together with the electron density. This indicates a depletion of neutrals ($O_2$, O, Ar) which could be due to a change in gas temperature, which is however not measurable, or to a change of the partial pressure given by the electrons. According to the data presented in fig. 3, the electron pressure ranges from about 9 % to 20 % of the total gas pressure. Being this pressure constant, with increasing RF power the electron pressure increases with a corresponding decrease of the pressure of the neutrals. According to the ionization balance model described in [14], this would in turn lead to an increase in electron temperature.

Another fact indicated by the data of fig. 3 is that for a given RF power both electron temperature and electron density decrease with increasing magnetic filter field. The maximum observed decreases, of the order of 10%÷15%, are well within the level of systematic errors; these should be however attributed to the absolute values of the measures, not to the relative variations between them, which are more accurate. The OES diagnostic has then detected and demonstrated the effectiveness of the magnetic filter field in NIO1.

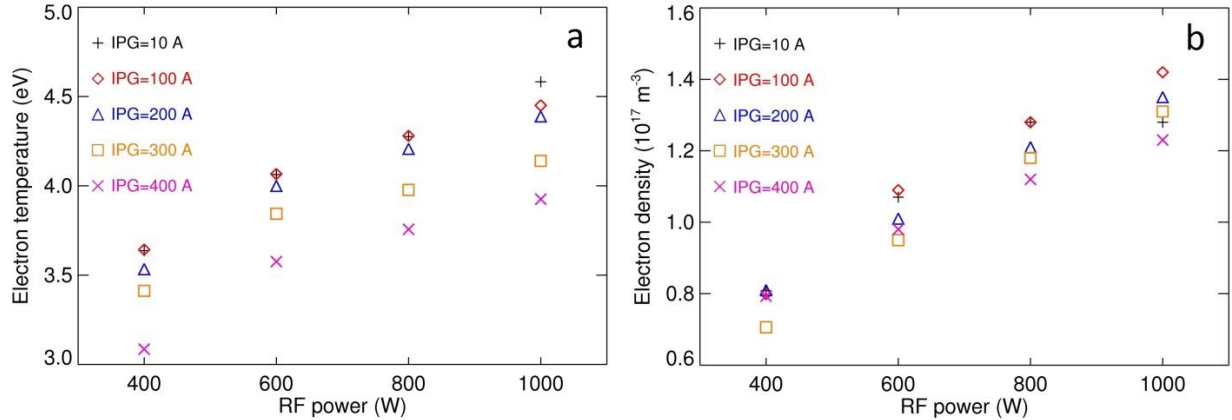

**FIGURE 3.** Measures of electron temperature (plot a) and electron density (plot b), obtained by scans in RF power performed at different values of PG current. The source pressure was 0.48 Pa; the PG current was brought back to the power supply through the "C" coils, while $BP_m$ and $BP_e$ were left at floating potential.

The results given by the OES diagnostic also confirmed the known phenomenon that an increase of the gas pressure leads to a decrease in electron temperature. This behavior is again foreseen by the model which describes the ionization balance in the plasma [14]: a higher pressure leads to a lower diffusion of ions towards the source walls; the ionization balance is then reached for a lower ionization reactivity, which is reached in $O_2$-Ar plasmas [15-16] at a lower electron temperature. This is shown in fig. 4a, where the electron temperature is plotted as function of the source pressure, for a plasma sustained by 1000 W RF power. $I_{PG}$ was set to 400 A, however the 2 data series belong to different magnetic configurations of the source. The points indicated with black diamonds were measured when the PG current circuit was closed on the "C" coils, while those indicated with red triangles were measured when the circuit was closed on the $BP_m$. In the latter case the magnetic field is stronger between $BP_m$ and PG (100 G/$I_{PG}$[kA]), but weaker at the LOS position with respect to the "C" coil configuration (25 G/$I_{PG}$[kA] instead of 75 G/$I_{PG}$[kA]) [4]. The electron temperature is then higher in the latter configuration, not only because of the weaker magnetic field but also because the electrons are slightly "pushed back" from the zone with higher magnetic field. It's also interesting to notice that the decrease of the electron temperature with the gas pressure is much less pronounced with the source configuration employing the $BP_m$. Moreover, the analysis of the cases shown in fig. 4a showed also that the electron density is essentially independent on the gas pressure (in the considered pressure range). From the measurements made with the "C" coil configuration an average electron density of $1.17 \cdot 10^{17}$ m$^{-3}$ (maximum statistical error ±7 %) resulted, while for the configuration using the $BP_m$ the average electron density was $1.4 \cdot 10^{17}$ m$^{-3}$ (maximum statistical error ±2 %). These values witness the different levels of magnetic filter field produced by the 2 source configurations at the LOS position.

At last, the OES diagnostic also detected the influence of the PG bias voltage on the properties of the electron gas. In fig. 4b the electron density is plotted as function of the bias voltage, which was applied between the PG and the source body; $BP_m$ and $BP_e$ were left at floating voltage. The shown measurements were performed on a plasma at 0.37 Pa pressure and sustained by 800 W RF power. No magnetic filter field was applied ($I_{PG}$=10 A), while the EG and the PA voltages were set respectively to 1.5 kV and 15 kV for beam extraction and acceleration. What results from fig 4b is that the electron density lowers with increasing bias voltage, probably because of the progressive attraction of the electrons towards the PG. From the same measurements of fig. 4b it resulted that the electron temperature is essentially independent of the bias voltage; in particular, the average value of the electron temperature measurements was 4.63 eV (maximum statistical error ±3%).

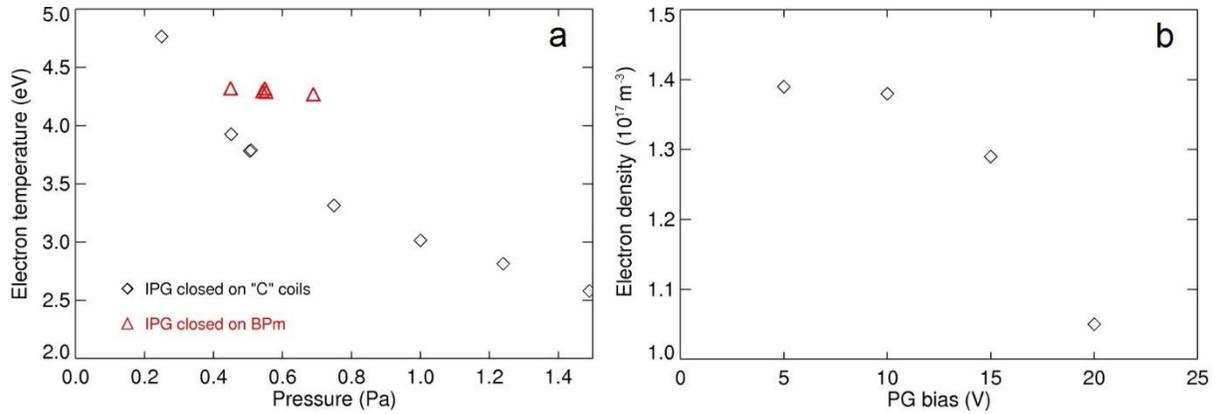

**FIGURE 4.** Plot a: Electron temperature measured as function of source pressure, for a plasma sustained by 1000 W RF power. In all the measurements the PG current was 400 A; the $I_{PG}$ circuit was closed on the external "C" coils for the points indicated with black diamonds, while it was closed on the $BP_m$ for the points indicated with red triangles. Plot b: Electron density measured as function of the bias voltage between PG and source body. The plasma was sustained by 800 W RF power at a pressure of 0.37 Pa; no magnetic filter field was present ($I_{PG}$=10 A), while EG and PA were set at 1.5 kV and 15 kV for beam extraction and acceleration.

## CONCLUSIONS

NIO1 has been been operated with a mixture of oxygen and argon to test the extraction system. Thanks to the OES diagnostic, some emission lines of argon were exploited to measure the electron temperature and the electron density in proximity of the Plasma Grid. The 2 quantities affect the production of negative ions and of (unwanted) co-extracted electrons. The interpretation of the data relied on the results of the ADAS CR model for Ar and $Ar^+$, which assumes a Maxwellian distribution for electrons; in the calculations it was also assumed that the gas temperature is 300°K, since no measurements techniques were available for the plasma conditions in NIO1 and with the available instrumentation.

The results given by the OES diagnostic and by the developed analysis method indicate that, for the typical regimes of pressures (~1 Pa) and RF power (~1kW) of NIO1, not only the electron density but also the electron temperature grow with increasing RF power. Both electron density and electron temperature were sensitive to the magnetic filter field, determined not only by the PG current but also by the different electric configurations adopted. The measurements also showed that, in NIO1 operative regime, an increase of gas pressure lowers the electron temperature, while leaving the electron density unchanged. On the contrary, an increase of PG bias voltage lowers the electron density close to the PG, but with no noticeable changes in the electron temperature.

## ACKNOWLEDGMENTS

This project has received funding from the European Union's Horizon 2020 research and innovation programme. The work leading to this publication has also been funded partially by Fusion for Energy under the Contract F4E-RFXPMS_A-WP-2015. This publication reflects the views only of the authors, and Fusion for Energy cannot be held responsible for any use which may be made of the information contained therein. The views and opinions expressed herein do not necessarily reflect those of the ITER Organization.